\documentclass[preprint, 3p]{elsarticle}

\pdfoutput=1 
\usepackage{import}
\usepackage{AMmath}
\usepackage[utf8x]{inputenc}
\usepackage[]{units}
\usepackage{ntheorem}

\usepackage{tikz}
\usepackage{pgfplots}
\AtBeginDocument{\catcode`\$=3} 
\usetikzlibrary{matrix}
\usepgfplotslibrary{groupplots}
\pgfplotsset{compat=newest}

\usepgfplotslibrary{external}
\newcommand{\R}[1]{\mathbb{R}^{#1}}

\newcommand{\Tau}{\mathcal{T}}
\renewcommand{\d}{\mathrm{d}}
\newtheorem{remark}{Remark}
\let\oldremark\remark
\renewcommand{\remark}{\oldremark\normalfont}

\renewcommand{\diff}[2]{\frac{\partial #1}{\partial #2}}

\begin{document}
\def\picwidth{0.8\linewidth}
\begin{frontmatter}
\title{Generalization of Quadratic Manifolds for Reduced Order Modeling of Nonlinear Structural Dynamics}

\author[tum]{J. B. Rutzmoser}
\author[tum]{D. J. Rixen\corref{daniel}}
\author[eth]{P. Tiso}
\author[eth]{S. Jain}

\cortext[daniel]{Corresponding author}
\address[tum]{
	Chair of Applied Mechanics, 
	Technial University of Munich, 
	D-85748 Garching, 
	Germany. \\
	Email: \{johannes.rutzmoser\}\{rixen\}@tum.de
}

\address[eth]{
	Institute for Mechanical Systems, 
	ETH Zürich, 
	Leonhardstrasse 21, 
	8092 Zürich, 
	Switzerland. \\
	Email: \{ptiso\}\{shjain\}@ethz.ch
}

\begin{abstract}
   In this paper, a generalization of a quadratic manifold approach for the reduction of geometrically nonlinear structural dynamics problems is presented. This generalization is constructed by a linearization of the static force with respect to the generalized coordinates, resulting in a shift of the quadratic behavior from the force to the manifold. In this framework, static derivatives emerge as natural extensions to modal derivatives for displacement fields other than the vibration modes, such as the Krylov subspace vectors. Here the dynamic problem is projected onto the tangent space of the quadratic manifold, allowing for a much less number of generalized coordinates compared to linear basis methods. The potential of the quadratic manifold approach is investigated in a numerical study, where several variations of the approach are compared on different examples, indicating a clear pattern where the proposed approach is applicable. 
	

\end{abstract}


\end{frontmatter}
\section{Introduction}
\label{sec:Introduction}
With the advent of the digital era and ever so powerful computers, elaborate simulation techniques have conquered the design processes in engineering. While the computational power has increased dramatically in the past decades, the demand for more detailed and accurate models still drives the development for even more efficient and powerful simulation techniques. 

In order to reduce the computational cost while preserving the accuracy of large scale models, Model Order Reduction (MOR) has found its way in the realm of nonlinear structural dynamics. Within the context of the Finite Element Method used for the spatial discretization of arbitrary geometries, two aspects need to be tackled together in order to solve the conflicting goals of low computational effort and high accuracy. The first aspect is the reduction of the number of unknowns in the governing equations based on subspace projection. These projective model order reduction techniques have seen huge success, especially for linear systems, as they allow for the reduction in the number of unknowns over several orders of magnitude while retaining the accuracy. The second aspect deals with the complexity reduction in the computation of certain nonlinear terms in the governing equations, which need to be updated for every evaluation in both, static and dynamic problems. In this aspect, usually referred to as Hyper Reduction, huge strides have been made recently in the context of Finite Elements \cite{farhat2015structure, tiso2013discrete}. 

In this paper, the first aspect, i.e. the reduction in the number of degrees of freedom (dofs) in the governing equations, is addressed. The key question of all of such projection-based techniques is about the subspace onto which the system of equations is projected. For linear systems, various systems-theory concepts exist, such as modal decomposition, observability and controlability, transfer function or linear superposition. Many reduction techniques rely on these properties to build a suitable reduction basis, e.g. Modal Truncation, Balanced Truncation, Krylov Subspace Methods with Moment Matching or other, rather physically-intuitive reduction techniques like the  Craig-Bampton Method. An overview of the mentioned methods with application to linear structural dynamics is given in \cite{witteveen2012modal}.
For nonlinear systems, however, these concepts either do not exist, or are not computationally feasible. For this reason, most nonlinear reduction techniques are not based on intrinsic physical properties upon which the linear methods rely, but rather are data driven, where an existing solution is analyzed in order to generate a subspace in which the solution is approximated. Despite the accuracy, all these techniques, mostly being a variant of the Proper Orthogonal Decomposition (POD), carry the drawback of the requirement of a full simulation in advance, which can seem to be against the goals of Model Order Reduction,  especially in cases where the available computational resources do not allow a full simulation. The need for MOR in such \textit{simulation free} scenarios has also been a motivation for this work.

In order to build a projection subspace for nonlinear structural dynamics independent of full \emph{a priori} simulations, some attempts have been made. The key strategy is the application of an established reduction scheme on the linearized model, followed by the extension of the reduction basis to capture the nonlinearity. A popular example is the use of Modal Derivatives (MDs), which are computed by means of perturbation of vibration modes \cite{idelsohn1985reduction, barbic2005real, witteveen2014efficient, tiso2016part_1}. Furthermore, the extension of the vibration modes to nonlinear systems with the so called nonlinear normal modes has also been used for MOR \cite{touze2008reduced-order}, or even for demonstrating the reduction quality of the MDs \cite{kuether2014evaluating, sombroek2015bridging}. 

As the computation of MDs is expensive and involves the solution of a singular system, usually a simplified \textit{static} version of the MDs is also used (cf. companion paper \cite{tiso2016part_1} for a discussion). As MDs or their simplified counterparts have proved to be efficient tools for Model order reduction \cite{idelsohn1985reduction, barbic2005real, weeger2015on-the-use-of-modal}, their concept has also been extended to other types of linear modes \cite{idelsohn1985load}, however lacking a sound theory. A further issue in augmenting a linear basis using MDs is the quadratic growth of the basis size with respect to the number of linear modes used initially. To tackle that issue and keep the reduction basis small, selection strategies for a specific augmentation were proposed in \cite{tiso2011optimal} and in the companion paper \cite{tiso2016part_1}. 

In this paper, the concept of these simplified Modal Derivatives is equipped with a sound theory by the use of a Quadratic Manifold (QM). The key idea lies in the mapping, not into a constant subspace, which can be interpreted as a linear, uncurved manifold, but into a Quadratic (nonlinear) Manifold. The projection subspace is then the position dependent tangent space of the Manifold. As already discussed in the companion paper \cite{tiso2016part_1}, the QM takes care issue of the quadratic growth in the linear basis size when the MDs are used as independent components by minimizing the number of unknowns for the given augmented linear basis. 

This paper complements the companion paper \cite{tiso2016part_1}, where the concept of QM is proposed, and tested on shell structures. Here, the theory of the Static MDs is extended to displacement fields other than the VMs, with a sound physical foundation. These are then called Static Derivatives (SDs). Furthermore, the proposed generalization is tested on a broader class of structures, namely 2D and 3D continuum Finite Element based structures. 

The paper is organized as follows. In the next section, after an introduction to linear projection for model order reduction, the concept of nonlinear projective model order reduction is presented using the QM approach as an example. Then, two strategies for the necessary ingredients for the QM are discussed, the approach using MDs and the \textit{Force Compensation Approach} using SDs. Thereafter, the use of SDs in the framework of a linear basis is discussed. Subsequently in Section \ref{sec:applications}, the proposed methods are applied to four examples with a focus on the accuracy of the given methods. In the last section the contribution of this paper is concluded. 

\section{Model Order Reduction using Quadratic Manifold}
\subsection{Linear Projective Model Order Reduction}
\label{sub:linear_projective_model_order_reduction}
The semi-discretized equations of motion of a (geometrically) nonlinear structure are given as
\begin{align}\label{eq:nonlinear_system}
\vM \ddot \vu + \vC\dot{\vu} + \vf(\vu) = \vg,
\end{align}
where $\vu\in\R{N}$ is the vector containing the physical displacement dofs, ~$\vM\in\R{N\times N}$ is the mass matrix,~$\vC\in\R{N\times N}$ the damping matrix,~$\vf(\vu)\in\R{N}$ is the nonlinear internal force vector, and $\vg\in\R{N}$, the external forcing. The time dependence of $\vu$ and $\vg$ is omitted for brevity. The basic idea of linear projective MOR for second order systems such as in ~\eqref{eq:nonlinear_system} is the linear transformation of the displacement vector~$\vu$ to a reduced set of generalized coordinates~$\vq\in\R{n}$, where $n\ll N$, such that 
\begin{align}\label{eq:linear_transformation}
	\vu = \vV\vq, \qquad \dot{\vu} = \vV\dot{\vq}, \qquad \ddot{\vu} = \vV\ddot{\vq}\,,
\end{align}
where $\vV\in\R{N\times n}$ is the projection matrix. The columns of $\vV$ span the subspace in which the high dimensional displacement vector~$\vu$ is constrained and the entries of $\vq$ represent the time varying amplitudes of the column vectors in $\vV$.

If \eqref{eq:linear_transformation} and its derivatives are substituted in \eqref{eq:nonlinear_system}, a residual $\vr$ is expected as the system of equations would not be satisfied in general. More specifically,
\begin{align}\label{eq:eq_of_motion_residual}
	\vM \vV\ddot \vq + \vC\vV\dot{\vq} + \vf(\vV\vq) = \vg + \vr.
\end{align}
According to the principle of virtual work, the residual force $\vr$ should be orthogonal to the kinematically admissible motion $\delta \vu = \vV\delta \vq$.  In order to solve the under-determined system \eqref{eq:eq_of_motion_residual}, Galerkin projection can be used, which implies $\vV^T\vr=\mathbf{0}$. This results in the reduced system of equations
\begin{align}\label{eq:reduced_equation_of_motion}
	\vV^T\vM \vV\ddot \vq + \vV^T\vC\vV\dot{\vq} + \vV^T\vf(\vV\vq) = \vV^T\vg,
\end{align}
or, equivalently,
\begin{equation}
	\vM_{r} \ddot\vq + \vC_r\dot{\vq} + \vf_r(\vq) = \vV^T\vg,
\end{equation}
where $\vM_r = \vV^T\vM\vV$ and $\vC_r = \vV^T\vC\vV$ are the reduced linear components which can be precomputed offline, and $\vf_r(\vq) = \vV^T\vf(\vV\vq)$ is the reduced nonlinear force. Clearly, this resulting system is of dimension~$n<<N$. 

\subsection{A Nonlinear Projective Model Order Reduction Approach: Quadratic Manifold}
\label{sub:quadratic_manifold}
The key idea in linear projective Model order reduction is the linear mapping between the full, high dimensional displacement vector~$\vu$ and the reduced variable~$\vq$. However, this mapping can be generalized to a nonlinear mapping $\vGamma(\vz):\mathbb{R}^n \mapsto \R{N}$, with $\vz\in\R{n}$ being the vector of nonlinear reduced generalized coordinates such that
\begin{align}\label{eq:nonlinear_transformation}
	\vu = \vGamma(\vz), \qquad \dot{\vu} = \frac{\partial \vGamma}{\partial \vz}\dot{\vz}, \qquad \ddot{\vu} = \frac{\partial \vGamma}{\partial \vz}\ddot{\vz} + \frac{\partial^2 \vGamma}{\partial \vz^2}\dot{\vz}\dot{\vz}.
\end{align}
The nonlinear transformation \eqref{eq:nonlinear_transformation} is then substituted into the governing equations \eqref{eq:nonlinear_system}, and the force residual~$\vr$ is chosen to be orthogonal to the kinematically admissible displacements~$\delta \vu$ according to the principle of virtual work. For this transformation, $\delta \vu$ is then given by
\begin{align}
	\delta \vu = \frac{\partial \vGamma}{\partial \vz}\;\delta \vz = \vP_{\Gamma}\;\delta\vz,
\end{align}
with the tangent projector $\vP_{\Gamma}(\vz)=\frac{\partial \vGamma}{\partial \vz}$ spanning the tangent subspace of the kinematically admissible displacements $\delta \vu$ . This results in the reduced system of equations as
\begin{align}\label{eq:nonlinear_reduced_equation_of_motion}
	\vP_{\Gamma}^T\vM\vP_{\Gamma}\ddot{\vz} + \vP_{\Gamma}^T\vM\frac{\partial^2 \vGamma}{\partial \vz^2}\dot{\vz}\dot{\vz} + \vP_{\Gamma}^T\vC\vP_{\Gamma}\dot{\vz} + \vP_{\Gamma}^T\vf\left(\vGamma(\vq)\right) = \vP_{\Gamma}^T\vg,
\end{align}
with the reduced mass matrix $\vP_{\Gamma}^T\vM\vP_{\Gamma}\in\R{n\times n}$, the reduced damping matrix $\vP_{\Gamma}^T\vC\vP_{\Gamma}\in\R{n\times n}$ and the reduced nonlinear force $\vP_{\Gamma}^T\vf(\vGamma(\vz))\in\R{n}$. Relative to the linear projective reduced system \eqref{eq:reduced_equation_of_motion}, the extra term $\vP_{\Gamma}^T\vM\frac{\partial^2 \vGamma}{\partial \vz^2}\dot{\vz}\dot{\vz}$ can be interpreted as a convective term due to the change of the basis, which is proportional to the curvature $\frac{\partial^2 \Gamma}{\partial \vz^2}$ of the nonlinear transformation $\vGamma(\vz)$ and to the square of the generalized velocities~$\dot{\vz}$.

In the linear projective MOR using the  transformation \eqref{eq:linear_transformation}, the critical issue lies in a proper selection of the matrix~$\vV$. Similarly, in the framework of the nonlinear transformation~\eqref{eq:nonlinear_transformation}, an appropriate non-linear function~$\vGamma(\vz)$ is sought. In this paper, a QM-based approach, as introduced in \cite{tiso2016part_1}, is used. The transformation $\vGamma(\vz)$ is a quadratic function in $\vz$, which can be cast in the form
\begin{align}\label{eq:quadratic_transformation}
	\vu = \vGamma(\vz) = \vV\vz + \frac{1}{2}\vTheta\vz\vz,
\end{align}
where the linear part $\vV\vz$ is formed by the matrix $\vV\in\R{N\times n}$, and the quadratic part $\vTheta\vz\vz$ is described using a symmetric\footnote{For a deeper discussion on the symmetry of $\vTheta$, the reader is referred to the companion paper \cite{tiso2016part_1}.} third order tensor $\vTheta\in\R{N\times n \times n}$. In this paper, the tensor-vector product is an inner product on the last index of the tensor similar to a matrix-vector product. Equation~\eqref{eq:quadratic_transformation} can be expressed, equivalently, in the indicial notation with implicit summation over repeated indices as
\[
u_i = V_{ij}z_j + \frac{1}{2}\theta_{ijk}z_kz_j.
\]
We denote the use of the transformation \eqref{eq:quadratic_transformation} for MOR as the Quadratic Manifold Approach. By substituting the mapping \eqref{eq:quadratic_transformation} into  \eqref{eq:nonlinear_transformation}, we obtain the displacement, velocity and acceleration vectors as 
\begin{gather}
	\vu = \vV\vz + \frac{1}{2} \vTheta \vz\vz, \quad \dot{\vu} = \vP_{\Gamma}\dot{\vz}, \quad \ddot{\vu} = \vP_{\Gamma} \ddot{\vz} + \vTheta\dot{\vz}\dot{\vz},	
\end{gather}
where the tangential projector~$\vP_{\Gamma}\in\R{N\times n}$ is given as
\begin{equation*}
\vP_{\Gamma} = \vV + \vTheta\vz.
\end{equation*}
Clearly, a proper selection of~$\vV$ as well as the quadratic transformation part~$\vTheta$ is critical in ensuring the success of this approach.
\subsection{Modal Derivatives}
\label{sub:modal_derivatives}
In the reduction of linear structural mechanical systems, the choice of the basis~$\vV$ as vibration modes is very common (cf. eg. \cite{geradin2014mechanical}). Therefore, if the nonlinear force $\vf(\vu)$ is linearized and can be cast in the form $\vK\vu$, one can evaluate the generalized eigenvalue problem
\begin{align}
	(-\omega_i^2\vM + \vK)\vphi_i = 0
\end{align}
associated with the linear equation of motion, yielding the $i$-th vibration mode~$\vphi_i$ with the corresponding undamped eigenfrequency $\omega_i$. As is common in linear reduction, the linear basis $\vV$ is composed of vibration modes such that $\vV = [\vphi_1, \vphi_2, \dots, \vphi_n ]$, one can fill the Quadratic Tensor $\vTheta$ with Modal Derivatives, as they describe the change of one mode with respect to another. The general idea behind MDs is the perturbation of the eigenvalue problem associated with a linear, parameter-dependent mechanical system along the parameter~$p$ as
\begin{gather}
	\frac{\partial}{\partial p}\left[(-\omega_i^2\vM + \vK)\vphi_i \right] = 0 \\
	\frac{\partial}{\partial p}\left[(-\omega_i^2\vM + \vK) \right] \vphi_i + (-\omega_i^2\vM + \vK) \frac{\partial \vphi_i}{\partial p} = 0, \label{eq:md_with_p}
\end{gather}
yielding the derivative $\frac{\partial \vphi_i}{\partial p}$ of the eigenvector~$\vphi_i$ with respect to the parameter~$p$. If the parameter $p$ is taken to be amplitude $ q_j $ associated to eigenmode $ \vphi_j $, we obtain 
\begin{gather}\label{eq:modal_derivative}
	(-\omega_i^2\vM + \vK(q_j = 0))\frac{\partial \vphi_i}{\partial q_j} = - \left.\frac{\partial}{\partial q_j} \left[-\omega_i^2\vM + \vK(q_j)\right]\right|_{q_j = 0} \vphi_i,	
\end{gather}
\begin{equation}\label{key}
\text{where} \quad \vK(q_j) = \left.\frac{\partial \vf(\vu)}{\partial \vu} \right|_{\vu = \vphi_j q_j}
\end{equation}
and $\frac{\partial \vphi_i}{\partial q_j}$ is the modal derivative. A static variant of \eqref{eq:modal_derivative} is obtained with neglection of inertia terms \cite{idelsohn1985reduction, barbic2005real} resulting in:
\begin{align}\label{eq:static_modal_derivative}
	\vK \frac{\partial \vphi_i}{\partial q_j} = - \frac{\partial \vK}{\partial q_j} \vphi_i.
\end{align}
In the literature, both concepts of modal derivatives, with and without inertia terms, are referred to as modal derivatives. As they do not carry the same properties and have also a different meaning, we use the term \emph{Modal Derivative} (MD) for the derivative computed based on the perturbation of the full eigenvalue problem as in \eqref{eq:md_with_p}, whereas the derivative computed according to \eqref{eq:static_modal_derivative} is referred to as \emph{Static Modal Derivative} (SMD). The reader is referred to the companion paper \cite{tiso2016part_1} for a detailed discussion about the (S)MDs.

\subsection{Force Compensation Approach: Static Derivatives and Static Modal Derivatives}
In the companion paper \cite{tiso2016part_1}, it is assumed that the basis~$\vV$, representing the linear part of the mapping, is composed of a truncated set of vibration modes of the linearized, undamped system, and the quadratic part with ~$\vTheta$ is chosen to consist of the Modal Derivatives or the Static Modal Derivatives (SMDs). In this paper, we will discuss the proper choice of the quadratic part when the linear part of the transformation described by $\vV$ is not formed by vibration modes, but by any linear reduction basis suited to the problem at hand. While in the companion paper \cite{tiso2016part_1} the use of a quadratic manifold with (S)MDs is motivated by the extension of linear modal superposition through modal perturbation, this paper starts from a different point of view and shows that a generalization of the QM can be developed by searching for a non-linear manifold on which quadratic forces are linearized.
The quadratic mapping \eqref{eq:quadratic_transformation} involves a linear part given by $\vV$ and a quadratic part given by $\vTheta$. In the companion paper \cite{tiso2016part_1}, we discussed ways to construct $ \vTheta $ using (S)MDs, when $ \vV $ is composed of the vibration modes. Here we motivate the same construction in an alternative and general manner.

This apparoach is based directly on the source of nonlinearity i.e. the nonlinear force $\vf(\vu)$. If $\vu$ is expressed using the quadratic mapping  \eqref{eq:quadratic_transformation}, then choose the nonlinear $\vTheta$  such that the quadratic part of the nonlinear force~$\vf(\vGamma(\vz))$ vanishes. Formally, we require that
\begin{align}\label{eq:force_second_derivative_zero}
	\left.\frac{\partial^2 \vf(\vGamma(\vz))}{\partial \vz^2}\right|_{\vz = \mathbf{0}} = \mathbf{0}.
\end{align}
This condition defines the quadratic manifold such that it compensates the quadratic part of the nonlinear force with respect to the generalized coordinate~$\vz$. Then, quadratic tensor $\vTheta$ can be derived from \eqref{eq:force_second_derivative_zero} in the following manner. 

After applying the chain rule as
\begin{align}\label{eq:chain_rule_1}
\left[\left(\diff{^2\vf}{\vu^2}\cdot\diff{\vu}{\vz}\right)\diff{\vu}{\vz} + \diff{\vf}{\vu}\diff{^2\vu}{\vz^2}\right]_{\vz=0} = \mathbf{0},
\end{align}
with the first and second derivative of the quadratic mapping given as
\begin{align}
	\left.\diff{\vu}{\vz}\right|_{\vz=0} = \vV, \qquad
	\left.\diff{^2\vu}{\vz^2}\right|_{\vz=0} = \vTheta\label{eq:ddx_dz_theta},
\end{align}
and the definition of the stiffness matrix~$\vK = \left.\frac{\partial \vf(\vu)}{\vu}\right|_{\vu=0}$, one obtains the equation for the third order tensor~$\vTheta$ as
\begin{align}
	\label{eq:chain_rule_2}
	\left(\left.\diff{^2\vf}{\vu^2}\right|_{\vu=0}\cdot\vV\right)\vV + \vK \cdot \vTheta = \mathbf{0}.
\end{align}
From \eqref{eq:chain_rule_2}, the tensor $\vTheta$ is uniquely defined if it is taken to be symmetric with respect to the last two indices, i.e. $\Theta_{ijk} = \Theta_{ikj}$ (as is the case here), and $\vK$ is not rank deficient. In other words, that for \textit{any} projection matrix $\vV$ which forms the linear part of the quadratic mapping \eqref{eq:quadratic_transformation}, there exists a unique quadratic tensor $\vTheta$ compensating the nonlinear force to meet condition \eqref{eq:force_second_derivative_zero}. It is important to note that the linear part does not have to be composed of vibration modes, but can be composed of any suitably chosen set of linearly independent vectors. 

To further analyze \eqref{eq:chain_rule_2} we recast it into the indicial notation as
\begin{align}\label{eq:theta_definition_index}
\left( \left.\diff{^2 f_k}{u_l \partial u_m}\right|_{\vu=0}\right) V_{li} V_{mj} + K_{kl}\theta_{lij} = 0,
\end{align}
$k,l,m$ $ \in \{1,\dots,N\} $ are the indices of the physical domain, and $i,j  \in \{1,\dots,n\} $ are those of the reduced domain. The partial derivative $\left.\diff{u_m}{q_j}\right|_{\vu=0} = V_{mj}$ and the linearization of the nonlinear force $\left.\diff{f_k}{u_l}\right|_{\vu=\mathbf{0}} = K_{kl}$. Substituting these  relations into \eqref{eq:theta_definition_index}, one obtains
\begin{align}
\diff{}{u_m} \left( \left.\diff{f_k}{u_l} \right) \diff{u_m}{q_j} \right|_{\vu=0} V_{li} + K_{kl}\theta_{lij} = 0\\
\label{eqn:SDind} \implies \diff{}{u_m} \left( K_{kl} \right) \diff{u_m}{q_j} V_{li} + K_{kl}\theta_{lij} = 0.
\end{align}
And using the chain rule, \eqref{eqn:SDind} can be simplified to
\begin{align}
	\diff{K_{kl}}{q_j} V_{li} + K_{kl}\theta_{lij} = 0,
\end{align}
which can be rewritten in matrix-vector Notation with $l$ being the index of column vectors and $k$ being the row-index of matrices as follows:
\begin{align}\label{eq:static_correction_derivative}
\diff{\vK}{q_j}\vv_i + \vK\vtheta_{ij} = 0 \quad \Leftrightarrow \quad \vK\vtheta_{ij}  = -\diff{\vK}{q_j}\vv_i.
\end{align}
The solution of \eqref{eq:static_correction_derivative}, yielding the Static Derivatives (SDs) $ \vtheta_{ij} $, is equivalent to the definition of the SMDs \eqref{eq:static_correction_derivative}, if $\vV$ is chosen to consists of vibration modes: $\vV = [\vphi_1, \vphi_2, \dots, \vphi_n]$.  This implies that the SMDs are obtained as a special case of the general framework described by the \textit{Force Compensation Approach} using \eqref{eq:force_second_derivative_zero}. 

\begin{figure}\centering
   \def\svgwidth{\picwidth}
   {\footnotesize
      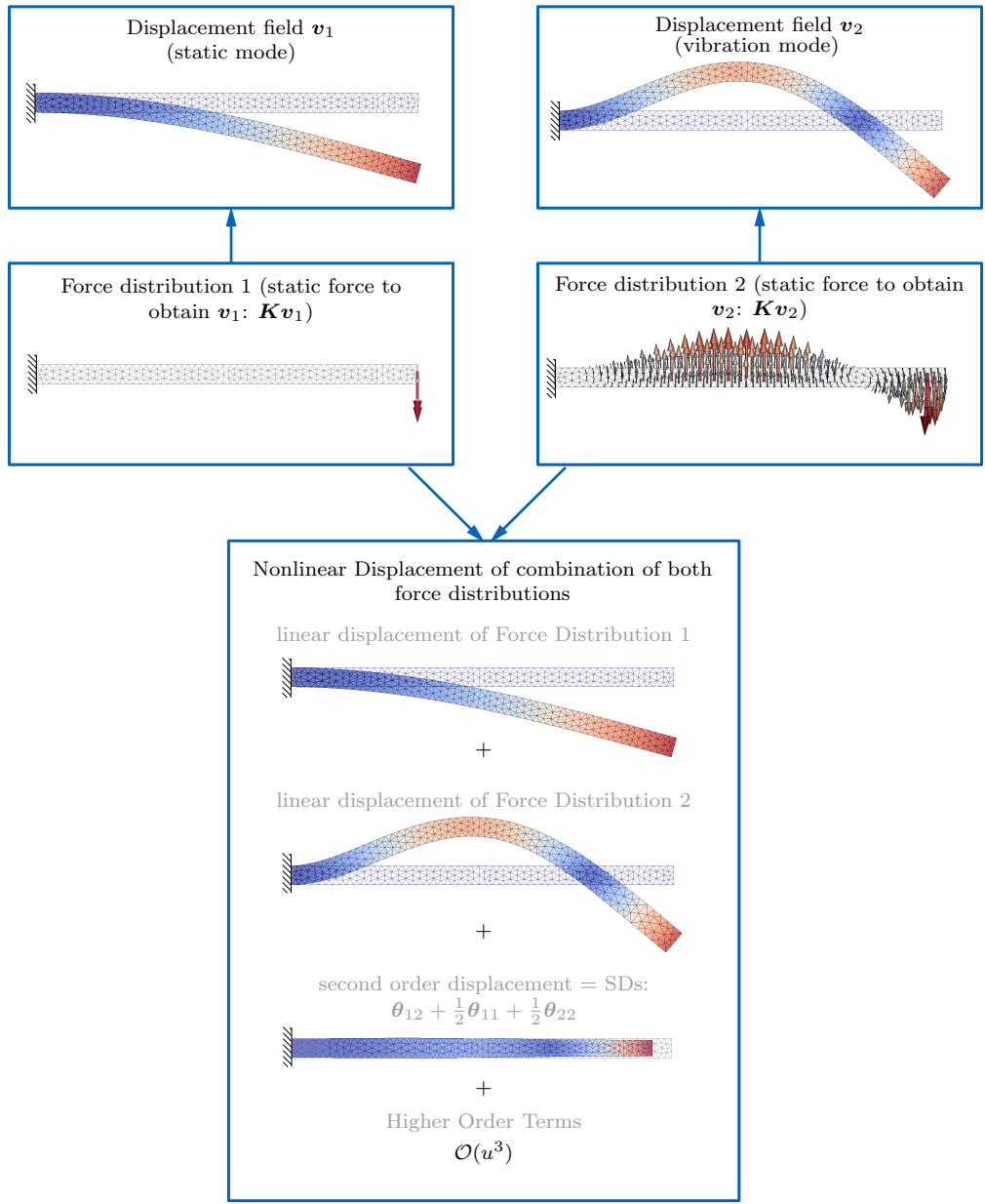}
   \caption{Physical interpretation of the static derivative (SD): Two force distributions yielding a linear displacement mode are applied to the nonlinear static problem. The result is a combination of the two linear displacement modes, the SDs in a second order expansion and higher order terms ($\mathcal{O}(u^3)$). \newline
   Note that for this beam example, both parent modes feature transverse displacements only, while the SD features only an axial field.}\label{fig:phys_interpretation_sds}
\end{figure}

It is emphasized that the approach presented above is not restricted to $ \vV $ consisting of vibration modes, but any set of vectors forming a suitable basis could be used, e.g. Krylov-Subspace vectors, static displacement fields or displacement vectors stemming from balancing controllability and observability.
The following remark gives a physical interpretation of the thus obtained SDs (cf. figure \ref{fig:phys_interpretation_sds}).

\begin{remark}[Physical Interpretation of Static Derivatives]

	For two given linear displacement fields $\vv_1$ and $\vv_2$ (e.g. $\vv_1$ being a static displacement mode of a unit force at the tip of the beam and $\vv_2$ being the second vibration mode), there exists a unique corresponding force distribution $\vf_1$ and $\vf_2$ such that $\vf_i = \vK\vv_i: i\in\{1,2\} $. If both force distributions are combined and applied to the \textit{nonlinear system}, the resulting displacement can be split in two contributions: a linear displacement $\vv_1 + \vv_2$, and a nonlinear correction resulting from the combination of both forces. As is shown below, the quadratic part of this nonlinear contribution contains \textit{exactly} all the SDs associated to the displacement fields $\vv_1$ and $\vv_2$.

	Under the assumption of analyticity, the nonlinear force can be extended using a Taylor expansion up to the quadratic part as
	\begin{align}\label{eq:nonlinear_force_taylor_expansion}
		\vf = \frac{\partial \vf}{\partial \vu} \vu + \frac{1}{2}\frac{\partial^2 \vf}{\partial \vu^2} \vu \vu + \mathcal{O}(u^3) = \vK \vu + \vK_2\vu\vu + \mathcal{O}(||\vu||^3)
	\end{align}
	with the linear stiffness matrix $\vK\in\R{N\times N}$ and the second order stiffness tensor $\vK_2\in\R{N \times N \times N}$. 
	The external force $\vg$ in figure \ref{fig:phys_interpretation_sds} is the linear force composed both displacements $\vv_1$ and $\vv_2$, i.e.
	\begin{align}
		\vg = \epsilon\vK(\vv_1 + \vv_2),
	\end{align}
	where $ \epsilon>0 $ is a load scaling factor.	Setting the static equilibrium of the external forces $ \vg $ with the Taylor expanded internal force \eqref{eq:nonlinear_force_taylor_expansion}, one obtains
	\begin{align}
		\vg & = \vf \\
		\label{eqn:subs} \implies \epsilon\vK(\vv_1 + \vv_2) & = \vK \vu + \frac{1}{2}\vK_2 \vu\vu + \mathcal{O}(||\vu||^3)\,, 
	\end{align}
	with the unknown displacement field $\vu$. For $ \epsilon $ small enough, we may assume that the solution can be expanded in $ \epsilon $ as 
	\begin{equation}\label{eqn:expansion}
	\mathbf{u} = \epsilon \vu_{(1)} + \epsilon^2 \vu_{(2)} + \dots\,.
	\end{equation}
	Substituting \eqref{eqn:expansion} into \eqref{eqn:subs}, and comparing coefficients of different powers in $ \epsilon $, we obtain the leading order coefficient to the solution $ \vu $ as
	\begin{align}
		\vu_{(1)} = \vv_1 + \vv_2,
	\end{align}
	and the second order approximation as
	\begin{align}\label{eqn:u2}
		\vu_{(2)} = & - \frac{1}{2}\vK^{-1}\vK_2 (\vv_1 + \vv_2) (\vv_1 + \vv_2)\,.		 		
	\end{align}
	Now, the SD $\vtheta_{ij}$ is defined according to \eqref{eq:theta_definition_index} as
	\begin{align}
	\vtheta_{ij} = - \vK^{-1} \vK_2 \vv_i \vv_j.
	\end{align}	
	Thus, by using the fact that the SDs are symmetric (i.e. $\vtheta_{12} = \vtheta_{21}$), $\vu_{(2)}$ from \eqref{eqn:u2} can be expressed in terms of SDs as
	\begin{equation}\label{eqn:SD}
	\vu_{(2)} = \frac{1}{2}(\vtheta_{11} + \vtheta_{12} + \vtheta_{21} + \vtheta_{22})= \vtheta_{12} + \frac{1}{2}(\vtheta_{11} + \vtheta_{22})
	\end{equation}
	 Consequently, the displacement field is composed of all three static derivatives, $\vtheta_{11}$, $\vtheta_{22}$ and $\vtheta_{12}$.

\end{remark}

As shown in figure \ref{fig:phys_interpretation_sds}, for a simple structure like a cantilevered beam, the second order nonlinear displacement of transverse displacements are axial displacements. This is due to the fact that the nonlinear response of transverse forces produces curved displacement trajectories, which are seen as the combination of the transverse linear modes with axial motion generating the curved trajectories. The separation of out-of-plane modes with corresponding in-plane SDs is possible for planar structures like beams or plates. It should be noted that the concept of SDs is independent of the underlying discretization of the problem, i.e. the SDs of a slender Beam modeled with 1D-Beam elements are equivalent to the same when discretized with solid triangular elements.

The computation of the SDs can be performed according to \eqref{eq:static_correction_derivative} using a finite difference scheme to compute the derivative of the stiffness matrix. Note, that the stiffness matrix has to be factorized only once and $\vtheta_{ij}$ can be solved via backward substitution. Furthermore, the derivative of the stiffness matrix needs to be computed for every $j \in\{1,\dots,n\}$. If a central finite difference scheme is used, $\vK$ has to be assembled $2n+1$ times (once with no perturbation and $2n$ times with $\vK(\pm \varepsilon \vv_j)$ for all $j$). The symmetry of the SD-Tensor can also be exploited, i.e. $\vTheta_{ijk} = \vTheta_{ikj}$ and thus only $\frac{n(n+1)}{2}$ SMDs need to be computed. A detailed proof for the symmetry of the SMDs is given in the companion paper \cite{tiso2016part_1}. This proof also holds for any type of linear base vector $\vv_i$ and consequently for the SDs as well.

\begin{remark}[Finite difference scheme for the computation of $\frac{\partial \vK}{\partial q_j}$]\label{remark:finite_difference_scheme}

The computation of partial derivative $\frac{\partial \vK}{\partial q_j}$ of the stiffness matrix with respect to the parameter~$q_j$ can be performed using a finite difference scheme. This practically useful as in most FE-codes, the  analytically computed tangential stiffness matrix is accessible, but the higher order derivatives are usually not.

The finite difference approximation to obtain sensitivities can be very sensitive with respect to the step width~$h$ \cite{van1998rigorous}. The convergence of SMDs w.r.t. $ h $ was only gained for central differences gives as
\begin{align}\label{eq:partial_derivative_central_differences}
\frac{\partial \vK}{\partial q_j} = \frac{\vK(\vv_j\cdot h) - \vK(-\vv_j\cdot h)}{2h},
\end{align}
but not for non-central difference schemes such as the forward or the backward finite differences. This is due to the fact, that the right hand side $\frac{\partial \vK}{\partial q_j}\vv_i$ of \eqref{eq:static_correction_derivative} represents a force, where the static derivative $\vtheta_{ij}$ is the displacement associated with this force. As this force is contaminated with numerical errors due to the finite difference scheme, the numerical error can be amplified severely if the corresponding displacement fields characterized by low stiffness (as is the case).

In Figure \ref{fig:phys_interpretation_sds} for instance, the SD is an axial displacement field. Consequently, the right hand side of equation \eqref{eq:static_correction_derivative} provides an axial compression force, leading to the displacement field of the SD shown. However, small numerical errors generate spurious forces in the transverse direction that can result in significant transverse displacement since the stiffness in the transverse direction is clearly much smaller than in the axial direction. 

As an alternative to the central finite difference scheme, a forward finite difference scheme with correction of the rigid body motion of every single element, as proposed in \cite{van1998rigorous}, could also be used.

\end{remark}

\begin{remark}[Orthogonalization of $\vTheta$ with respect to $\vV$]\label{rem:orthogonalization_theta}
	As the tangent projector $\vP_{\Gamma} = \vV + \vTheta\vz$ of the QM-approach is a combination of both, the linear mapping contributions stored in $\vV$ and the quadratic contributions in $\vTheta$, $\vP_{\Gamma}$ could become rank deficient in general, which is clearly an issue during time integration of the reduced system. To avoid this issue, the following Gram-Schmidt-like orthogonalization scheme of $\vTheta$ with respect to $\vV$ is proposed.
	\begin{align}\label{eq:theta_perp_v}
		\vTheta_{\perp\vV} = (\vI - \sum_{i=1}^{n}\vv_i\vv_i^T)\vTheta, \quad \text{with} \quad \vv_i^T\vv_i = 1
	\end{align}
	Here, all components of $\vV$ are subtracted from $\vTheta$ leading to $\vTheta_{\perp\vV}$ with $\vv_i$ being the normalized $i^{\mathrm{th}}$ column of~$\vV$.  By construction, it holds that
	\begin{align}\label{eq:condition_theta_perp_v}
		\vV^T\vTheta_{\perp\vV} = \mathbf{0}.
	\end{align}
	Note, that for the tensor $\vTheta_{\perp\vV}$ the condition \eqref{eq:force_second_derivative_zero} is broken. Consider e.g. the extreme case of $\vV$ being square, $\vTheta_{\perp\vV}$ is forced to be identically zero. Consequently, a QM composed of $\vTheta_{\perp\vV}$ degenerates to a linear manifold in this limit case. However, since usually $ \vV $ is a very thin matrix, this issue expected to be of a much milder consequence.
\end{remark}

\subsection{Augmentation of Linear Basis}
\label{sub:augmentation_of_linear_basis}
MDs, SMDs (cf. \cite{tiso2016part_1}) or the generalized SDs (as proposed in this work), can all be used as the quadratic terms in the framework of QM. However, they can also be used for the augmentation of a linear projection basis as in \eqref{eq:linear_transformation} and \eqref{eq:reduced_equation_of_motion}. Since the MDs and SDs are not guaranteed to be linearly independent with respect to the linear projection basis, a deflation technique is necessary to recover the subspace spanned by $\vV$ and $\vTheta$ combined.

A commonly used deflation technique is with the use of Singular Value Decomposition (SVD). Using this technique, the column vectors of both, the linear basis and the quadratic part, (MDs, SMDs or SDs) are stacked in a matrix $\vR$ as
\begin{align}
	\vR =  [\vv_1, \dots, \vv_n, \vtheta_{11}, \dots, \vtheta_{1n}, \vtheta_{22}, \dots, \vtheta_{2n}, \dots, \dots, \vtheta_{nn}],
\end{align}
 onto which SVD is then applied, revealing the continuous rank decay of~$\vR$. More specifically, the SVD results in
\begin{equation}
	\vR =  \vU_{\text{svd}}\vSigma_{\text{svd}}\vV_{\text{svd}}^T,
\end{equation}
where $ \vU_{\text{svd}} $ and $ \vV_{\text{svd}} $ are orthonormal matrices containing the left and the right singular vectors respectively, and $ \vSigma_{\text{svd}} $ is a (rectangular) diagonal matrix containing the singular values. An orthogonal linear basis of rank $ m $ can then formed by choosing the $m$ left singular vectors associated with the $m$ largest singular values. One possibe manner in which $ m $ can be chosen is as follows:
\begin{align}\label{eq:deflation_condition}
		\vV = [\vu_1, \dots, \vu_m], \quad \text{with} \quad \rho\;\sigma_{\text{1}} > \sigma_{m+1},
\end{align}
where $ \sigma_1 > \sigma_2 >\dots $ denote the singular values arranged in a decreasing order, $\rho$ is a user defined small tolerance. In this work, $\rho$ is chosen to be $10^{-8}$. A numerically stable orthonormal basis~$\vV$ well suited for projective model order reduction is obtained after deflation.

\section{Applications}
\label{sec:applications}
The feasibility as well as the restrictions of the proposed methods are discussed here by means of the following examples. The focus is put on the accuracy and stability of the proposed MOR methods. In order to obtain significant speedups, the proposed methods have to be combined with Hyper Reduction techniques, as mentioned in the introduction. Since hyper-reduction was not the key focus of this work, the computational time in the different examples has not been reported.

In order to assess the accuracy of the reduced solutions, the following global error measure $GRE_M$ is used
\begin{align}
GRE_M = & \frac{\sqrt{\sum\limits_{t\in\Tau}\Delta \vu(t)^T \vM \Delta \vu(t)}}{\sqrt{ \sum\limits_{t\in\Tau} \vu_{\text{ref}}(t)^T \vM \vu_{\text{ref}}(t)}},  \label{eq:err_global}
\quad \text{with} \quad \Delta \vu(t) = \vu(t) - \vu_{\text{ref}}(t),
\end{align}
where $\Delta \vu(t)$ is the difference between the displacements $\vu(t)$ of the reduced model, and $\vu_{\text{ref}}(t)$ of the reference model and $\Tau = [0, \Delta t_{\text{save}}, 2\Delta t_{\text{save}}, \dots, t_{\text{end}}]$ is the set of all timesteps at which the results are saved.

\subsection{Approach to investigation of the proposed methods}
\label{sub:approach_to_investigation}

To inquire the proposed methods and compare them to the full solution, three groups of reduction techniques are created.
\begin{itemize}
	\item \textit{QM}: In this category, we consider reduction solely by means of the Quadratic Manifold approach described in subsection \ref{sub:quadratic_manifold}. These experiments are labeled with \emph{QM}.
	\item \emph{QM\_\text{orth}}: As the stability of the time integration scheme is insufficient in some cases, an orthogonalization of the SDs is performed with respect to the linear basis vectors in $\vV$ (cf. remark~\ref{rem:orthogonalization_theta}). These reduced models with orthogonalized third-order tensors form the second group of experiments and are labeled with \emph{QM\_\text{orth}}. 	
	\item \emph{LB}: The reduction using a linear basis approach, described in subsection \ref{sub:augmentation_of_linear_basis}, forms the third category.  In these cases, a subspace spanned by the vectors in $\vV$ and $\vTheta$ is  used for reduction. A deflation is applied in order to keep the combined basis orthogonal. This group of experiments is labeled with \emph{LB}. 
\end{itemize}

The reduced models obtained by the LB and the QM approach result in a different number of reduced dofs for the same number of linear modes initially chosen. In the QM approach, the number of dofs is equal to the number of modes chosen \textit{a priori}. In the LB approach, the number of dofs is the number of linearly independent vectors in both, $\vV$ and $\vTheta$, as identified by the deflation technique (cf. subsection \ref{sub:augmentation_of_linear_basis}). Since we compare reduced order models resulting from the same number of linear modes, the number of reduced dofs varies between QM-reduced and LB-reduced models. The numerical experiments are organized as follows:

\begin{itemize}
	\item linearized: The full system is linearized around the point of equilibrium. This simulation gives a reference to assess the significance of nonlinear effects present in the nonlinear simulations.
	\item \textit{QM}
	\begin{itemize}
	\item QM-MD: Quadratic Manifold approach with linear Vibration Modes and Modal Derivatives as given by \eqref{eq:modal_derivative}. 
	\item QM-SMD: Quadratic manifold approach with linear Vibration Modes and Static Modal Derivatives (SMDs) as given in equation \eqref{eq:static_modal_derivative}.
	\item QM-KrySD: Quadratic manifold approach using the force compensation approach where Krylov-modes form the linear basis. The Krylov modes stem from the sequence
	\begin{align}
	\vV_{\text{kry, raw}} = [\vK^{-1}\vF, \vK^{-1}\vM\vK^{-1}\vF, \dots, (\vK^{-1}\vM)^{n-1}\vK^{-1}\vF],
	\end{align}
	which are $\vM$-orthogonalized using a Lanczos iteration to gain $\vV_{\text{kry}}$ \cite{wilson1982dynamic}. The basis $\vV_{\text{kry}}$ matches the $n$ moments around the static configuration, i.e. for the zero-frequency $\omega=0$. The $\vTheta$ is then formed with the SDs (given by \eqref{eq:static_correction_derivative}) obtained from the vectors in $\vV_{\text{kry}}$.
	\item QM-KrySD-SMD: Quadratic manifold approach where the linear basis is composed of both, Krylov modes around $\omega=0$ and vibration modes using the static compensation method. Half of the linear basis is composed of the Krylov vectors, and the other half is composed of the vibration modes. The basis $\vV$, formed by collecting both these contributions, is then orthogonalized to avoid bad conditioning. The $\vTheta$ is again formed by the Force Compensation Approach (given by \eqref{eq:static_correction_derivative}) applied to the vectors in $\vV$.
	\end{itemize}
	\item \textit{QM\_orth}
	\begin{itemize}
	 \item QM-SMD-orth: Same as QM-SMD, but the column vectors in $\vTheta$ are orthogonalized with respect to $\vV$ as in \eqref{eq:theta_perp_v}.
	\item QM-KrySD-orth: Same as in QM-KrySD, but also with orthogonalization of $\vTheta$ with respect to $\vV_{\text{kry}}$.
	\item QM-Kry-SMD-orth: Same as in QM-KrySD-SMD but also with orthogonalization of $\vTheta$ with respect to $\vV$.
	\end{itemize}
	\item \textit{LB}
	\begin{itemize}
	 \item LB-MD: Linear basis based on vibration modes and MDs. The deflation is performed according to eq. \eqref{eq:deflation_condition}.
	\item LB-SMD: Linear basis based on vibration modes and SMDs. Deflation is performed according to eq. \eqref{eq:deflation_condition}.
	\item LB-KrySD: Linear basis based on Krylov-Subspace modes and their corresponding static derivatives. The deflation is performed according to eq. \eqref{eq:deflation_condition}.
	\item LB-KrySD-SMD: Linear basis composed by an equal number of Krylov-Subspace-modes and vibration modes, with their corresponding SDs. The deflation is performed according to eq. \eqref{eq:deflation_condition}.
	\end{itemize}
\end{itemize}

\subsection{Clamped-clamped beam}

First, we consider a clamped-clamped bending problem, inspired by the simple plate bending problem of the companion paper \cite{tiso2016part_1}. 
There, the plate is modeled using shell elements incorporating the von Karman kinematic assumption. For the sake of generality, here the beam shape is discretized using continuum elements with quadratic shape functions, and no specific kinematic assumptions have been adopted.

\begin{figure}[htb]\centering
   \def\svgwidth{\picwidth}
   {\footnotesize
      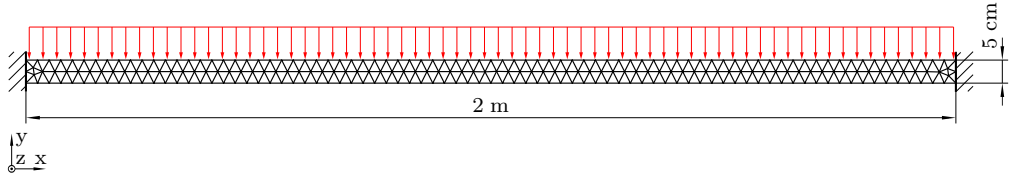}
   \caption{Mesh of the example Beam clamped-clamped with Dirichlet and Neumann Boundary conditions. The beam is exposed to a uniform vertical load on the top side. }\label{fig:benchmark_beam_cc}
\end{figure}

The mesh of the clamped-clamped beam example is depicted in figure \ref{fig:benchmark_beam_cc}. The beam is \unit[2]{m} long, \unit[5]{cm} high and clamped on both sides. The material model is Kirchhoff linear-elastic model, with a Young's modulus $E = \unit[70]{GPa}$, Poisson's ratio $\nu=0.3$ and density $\rho=\unitfrac[2700]{kg}{m^3}$.

\begin{figure}[htb]\centering
   \includegraphics[width=\picwidth]{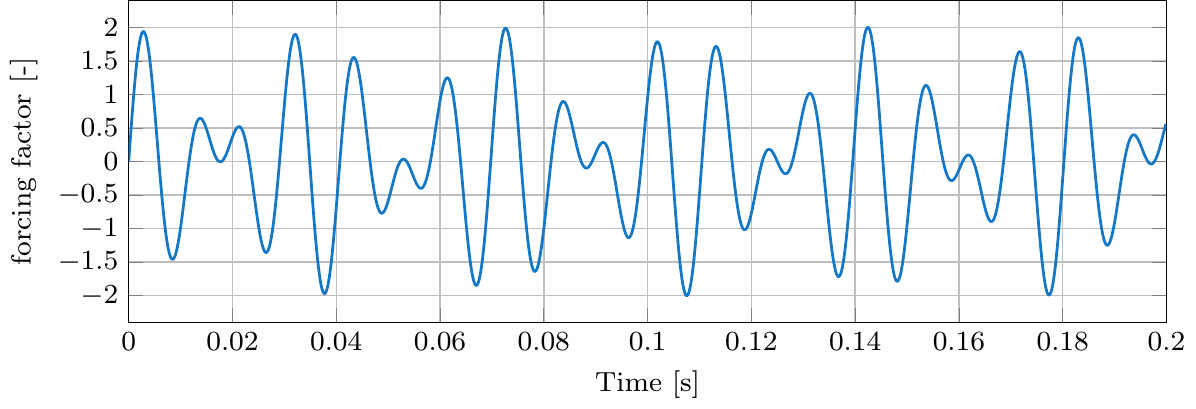}
   \caption{Time evolution of the forcing function of the Beam clamped-clamped.} \label{fig:forcing_time_series_clamped_clamped}
\end{figure}

The beam is uniformly loaded on the top with a force per unit length of $\unitfrac[5\cdot10^5]{N}{m}$ and the time dependent forcing function is $g(t) = \sin(72\cdot2\pi t) + \sin(100\cdot2\pi t))$, as shown in figure~\ref{fig:forcing_time_series_clamped_clamped}. The first three eigenfrequencies of the linearized structure are $\omega_1 = \unitfrac[65.2\cdot2\pi]{rad}{s}$, $\omega_2 = \unitfrac[178.8\cdot2\pi]{rad}{s}$ and $\omega_3 = \unitfrac[348.0\cdot2\pi]{rad}{s}$. The beam is discretized using $326$ quadratic triangular elements and has $1614$ dofs.

The time integration is performed using the HHT-$\alpha$ integration scheme with numerical damping $\alpha=0.1$ and time step size $\Delta t = \unit[1\cdot10^{-4}]{s}$. The integration was performed until $t_{\text{end}}=\unit[0.2]{s}$ while the displacement was saved every $\Delta t_{\text{save}} = \unit[1\cdot10^{-4}]{s}$.

\begin{figure}[htb]\centering
   \includegraphics[width=\picwidth]{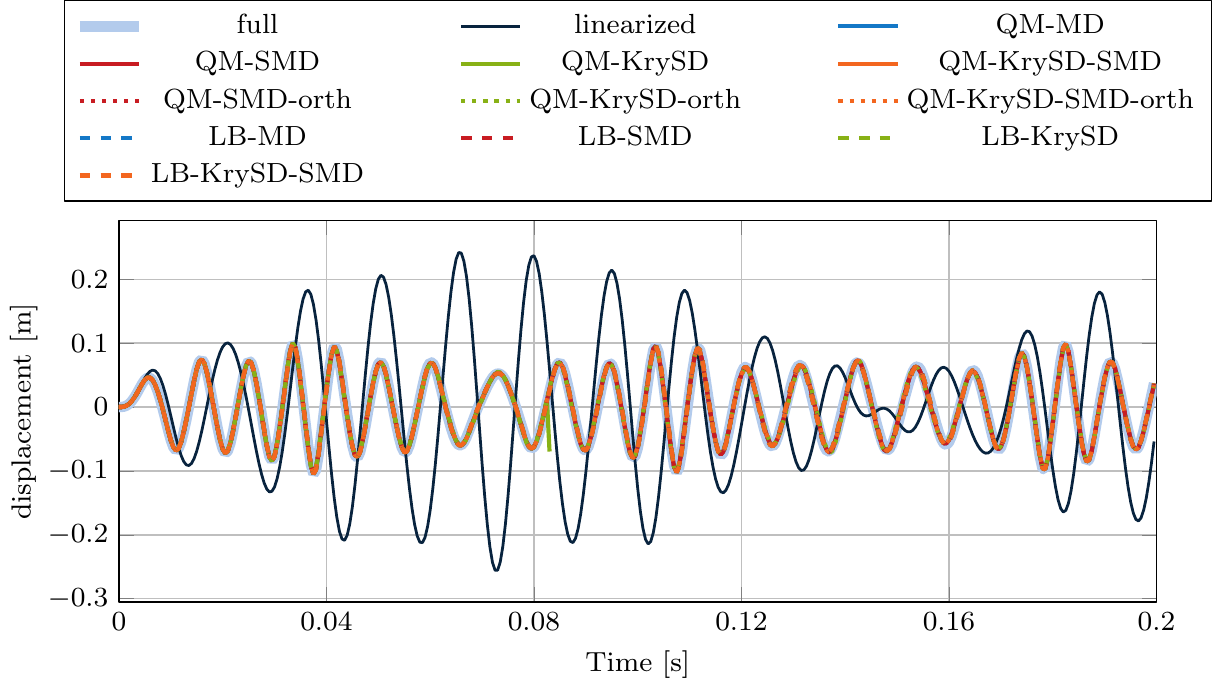}
   
   \caption{Time evolution of the displacement of a point in the middle of the clamped-clamped beam. The reduction methods are run with 10 modes.} \label{fig:time_evolution_beam_cc}
\end{figure}

The displacement in the transverse direction of a node situated in the middle of the beam is compared for the different reduction techniques listed in subsection \ref{sub:approach_to_investigation}. This comparison, using 10 modes in the linear basis, is depicted in figure~\ref{fig:time_evolution_beam_cc}. The linearized analysis highlights the significance of the nonlinear effects. The displacements obtained with all reduction methods show a good correlation with the full run.

\begin{figure}[htb]\centering
   \includegraphics[width=\picwidth]{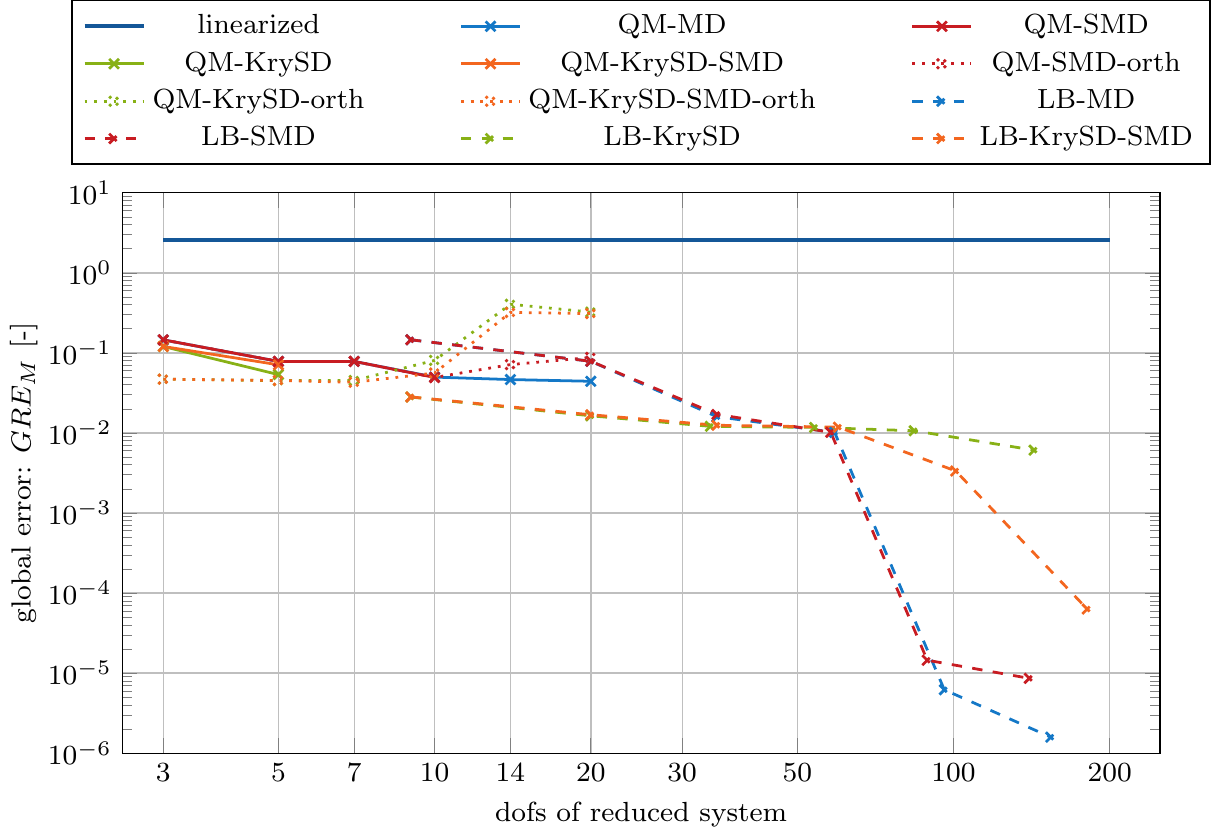}
   \caption{$GRE_M$ error of the benchmark example Beam clamped-clamped for the different reduction techniques. } \label{fig:error_overview_beam_cc}
\end{figure}

A closer look at the $GRE_M$ error for different reduction rates is given in figure~\ref{fig:error_overview_beam_cc}. For a low number of dofs, the accuracy of the QM methods is of the same order of magnitude as with the other reduction methods. Within the different techniques, the orthogonalized Krylov-methods show the best accuracy. Upon increasing the number of dofs in the reduced system, the QM approach tends to become unstable, with one exception, the QM-MD approach. Furthermore, the orthogonalization of the quadratic part (cf. Remark~\ref{rem:orthogonalization_theta}) solves this issue of stability, however with a loss of accuracy for an increasing number of modes. This is presumably due to the fact, that the quadratic relationship of modes and the corresponding (S)MDs gets violated with the procedure of orthogonalization. The LB method is also in the same range of accuracy as the QM methods up to 40 reduced dofs. For a higher number of dofs, especially the LB methods where (S)MDs are used, a gain in accuracy is observable.

\subsection{Hat}
\begin{figure}[htb]\centering
   \def\svgwidth{1.0\linewidth}
   {\scriptsize
      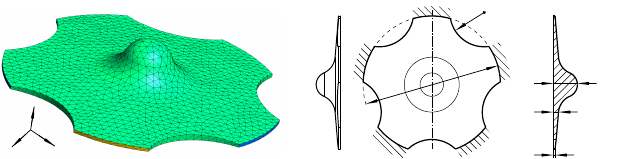}
   \caption{Mesh and dimensions of the benchmark example Hat. The outer rim surface (colored) is fixed, the bottom surface (not visible) is subjected to a uniformly distributed time dependent pressure. The mesh is composed of quadratic tetrahedron elements.}\label{fig:benchmark_hat}
\end{figure}

The second example which is investigated is a hat-shaped body. Sketches of its geometry and its mesh are depicted in figure~\ref{fig:benchmark_hat}. As in the previous example, a Kirchhoff material with Young's modulus  $E = \unit[70]{GPa}$, Poisson's ratio $\nu=0.3$ and density $\rho=\unitfrac[2700]{kg}{m^3}$ is adopted. The time integration is performed using a HHT-$\alpha$ integration scheme with $\alpha=0.1$, a time step size for the time integration is $\Delta t = \unit[4\cdot10^{-6}]{s}$, a time span from zero to $t_{\text{end}} = \unit[0.01]{s}$ and an export time step $\Delta t_{\text{save}} = \unit[4\cdot10^{-5}]{s}$.

\begin{figure}[htb]\centering
   \includegraphics[width=\picwidth]{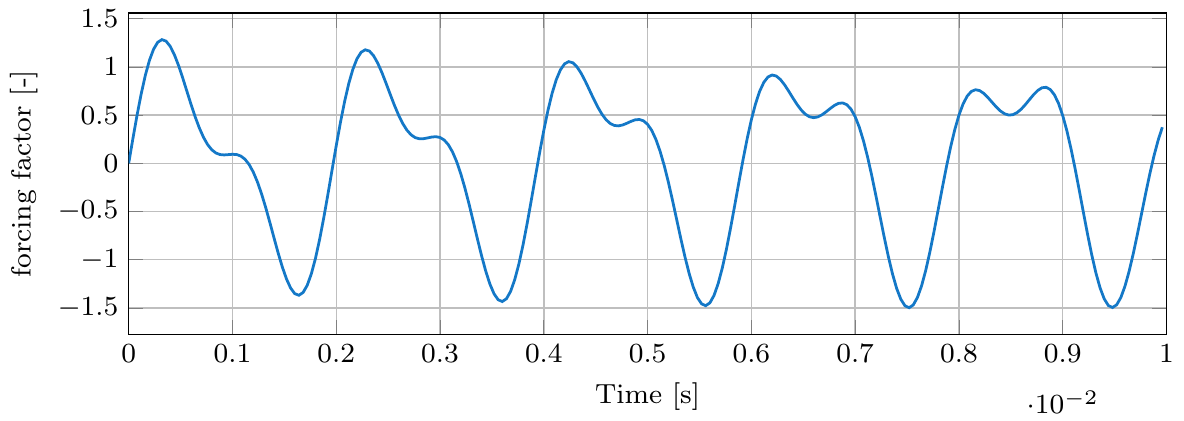}
   \caption{Time evolution of the forcing function of the hat example.} \label{fig:forcing_time_series_hat}
\end{figure}

The mesh is composed of 9774 tetrahedron elements having 52362 dofs in total. The part is fixed on the cylindrical outer surface and loaded with a constantly distributed vertical pressure in $y$-direction at the bottom with amplitude $\unitfrac[1\cdot10^7]{N}{m^2}$ and the time series $g(t) = \sin(500\cdot2\pi t) + \frac{1}{2}\sin(1030\cdot2\pi t))$, as shown in figure \ref{fig:forcing_time_series_hat}. The first eigenfrequencies of the linearized structure are $\omega_1 = \unitfrac[467.3\cdot2\pi]{rad}{s}$, $\omega_2 = \unitfrac[929.7\cdot2\pi]{rad}{s}$ and $\omega_3 = \unitfrac[930.1\cdot2\pi]{rad}{s}$.

\begin{figure}[htb]\centering
   \includegraphics[width=\picwidth]{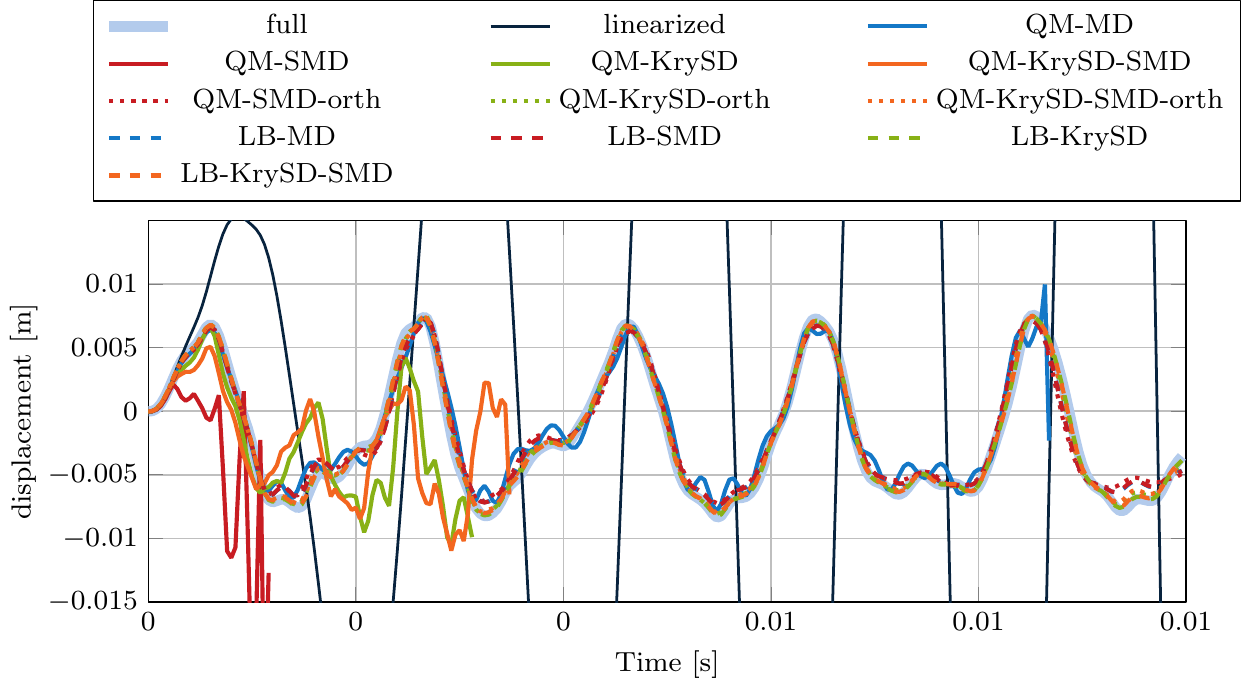}
   \caption{Time evolution of the displacement of a point in the middle of the hat. All reduction methods are run with 10 modes.} \label{fig:time_evolution_hat}
\end{figure}

The time evolution of the $y$-displacement of a node in the middle of the bottom surface is depicted in figure~\ref{fig:time_evolution_hat}. Clearly, the system is in the nonlinear regime as the displacements of the linearized system are extremely large, when compared to the reference solution. Furthermore, some of the QM approaches fail to converge, while the orthogonalized QM and the LB show a good agreement with the reference solution.

\begin{figure}[htb]\centering
   \includegraphics[width=\picwidth]{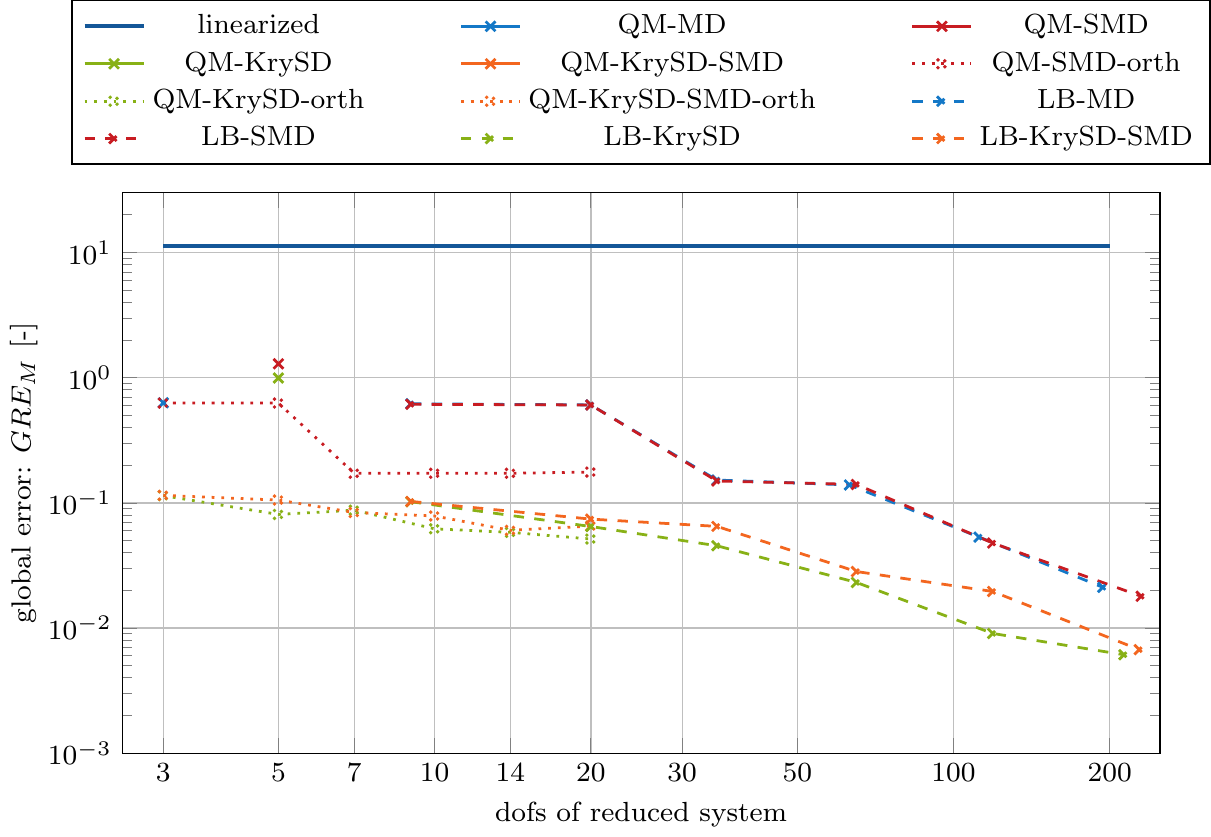}
   \caption{$GRE_M$ error of the example Hat for a different number of dofs and the proposed different reduction schemes.} \label{fig:error_overview_hat}
\end{figure}

A closer look at the error study with several reduction orders (cf. figure~ \ref{fig:error_overview_hat}) shows a trend similar to the Beam-cc example above. The QM approach without orthogonalization shows instability in most cases, but orthogonalization of the quadratic part solves this issue. Contrary to the clamped-clamped beam before, the orthogonalization does not lead to a loss in accuracy when the number of dofs is increased up to 20. However, this effect can be expected to set in with higher numbers of modes. The LB approach is in the same range of accuracy as the QM approaches for a small number of dofs, but gains accuracy when the number of modes is increased.

In both cases, the QM and the LB approach, the Krylov basis performs better compared to the modal approaches. This is presumably due to the fact, that the structure is symmetric and also symmetrically loaded. While the vibration modes contain several anti-symmetric modes which will be triggered barely for this load case, the Krylov subspace technique takes into account both, the symmetric structure and loading leading to basis vectors accounting for the symmetry.

\subsection{Arch}
\begin{figure}[htb]\centering
   \def\svgwidth{\picwidth}
   {\footnotesize
      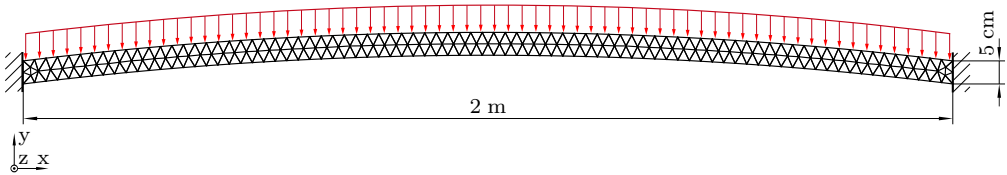}
   \caption{Mesh of the example Arch, which is clamped on both sides. }\label{fig:benchmark_arch}
\end{figure}

\begin{figure}[htb]\centering
   \includegraphics[width=\picwidth]{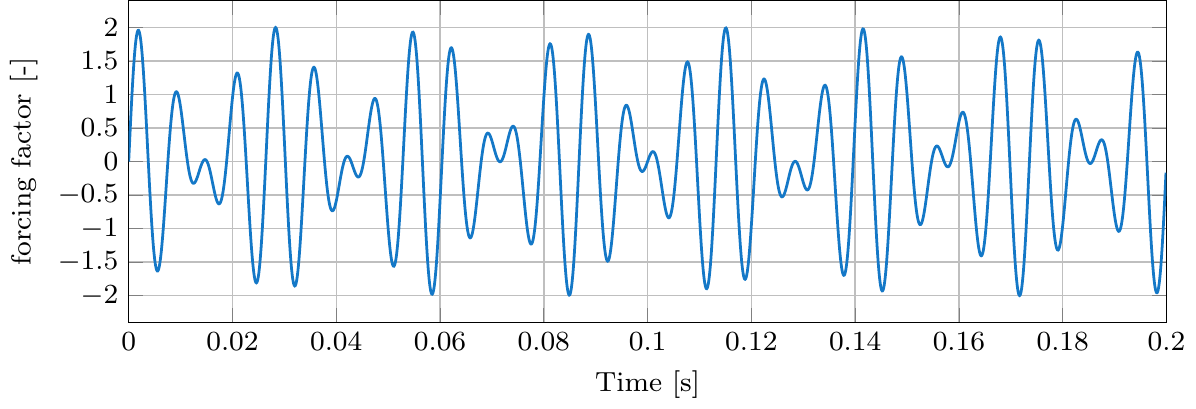}
   \caption{Time evolution of the forcing function of the Arch example.} \label{fig:forcing_time_series_arch}
\end{figure}

The next example is a clamped-clamped arch structure depicted in figure \ref{fig:benchmark_arch}. The arch is \unit[2]{m} long and \unit[5]{cm} high and has an arch radius of \unit[8]{m}. The geometry is discretized with 330 triangular elements with quadratic shape functions and has 1634 dofs in total. The adopted material is the same as in the first clamped-clamped beam example. The structure is loaded with a constant pressure in negative $y$-direction with amplitude $\unitfrac[3\cdot10^5]{N}{m}$, which is amplified with the loading time series $g(t) = \sin(115\cdot2\pi t) + \sin(150\cdot2\pi t))$ also depicted in figure \ref{fig:forcing_time_series_arch}. The first eigenfrequencies of the linearized structure are $\omega_1 = \unitfrac[104.6\cdot2\pi]{rad}{s}$, $\omega_2 = \unitfrac[176.2\cdot2\pi]{rad}{s}$ and $\omega_3 = \unitfrac[345.5\cdot2\pi]{rad}{s}$.

The time integration is performed with identical parameters as the beam clamped-clamped example.

\begin{figure}[htb]\centering
   \includegraphics[width=\picwidth]{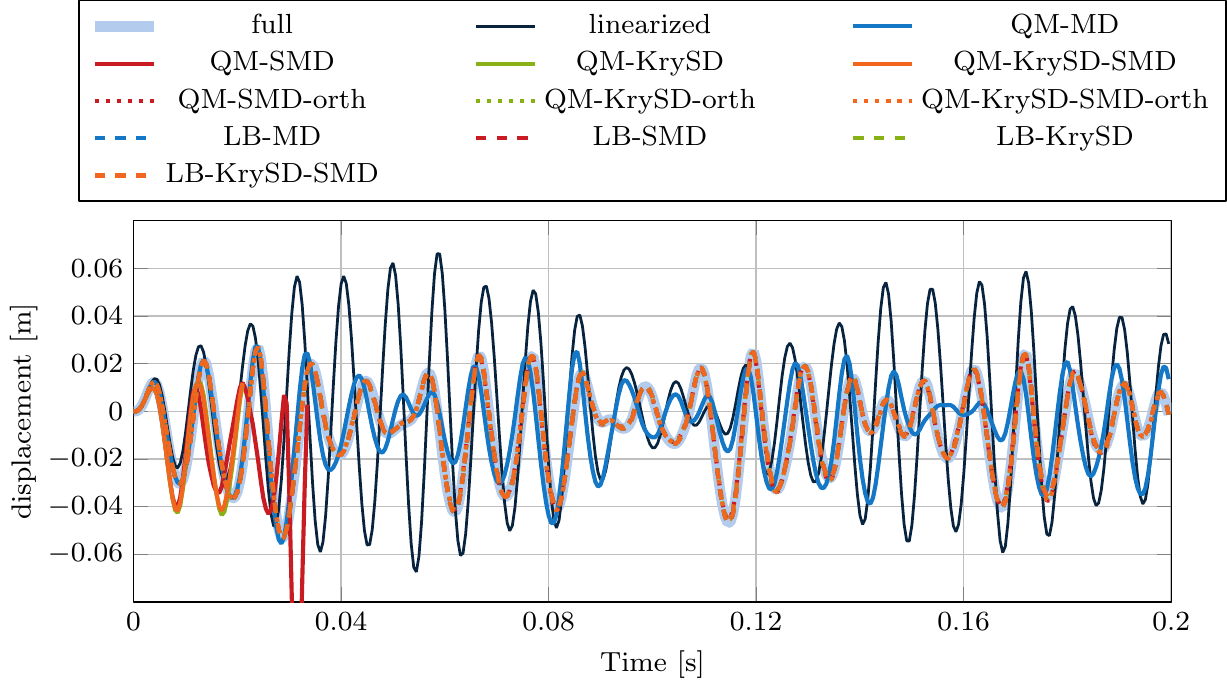}
   \caption{Time evolution of the displacement of a point in the middle of the arch. The reduction methods are run with 10 modes.} \label{fig:time_evolution_arch}
\end{figure}

\begin{figure}[htb]\centering
   \includegraphics[width=\picwidth]{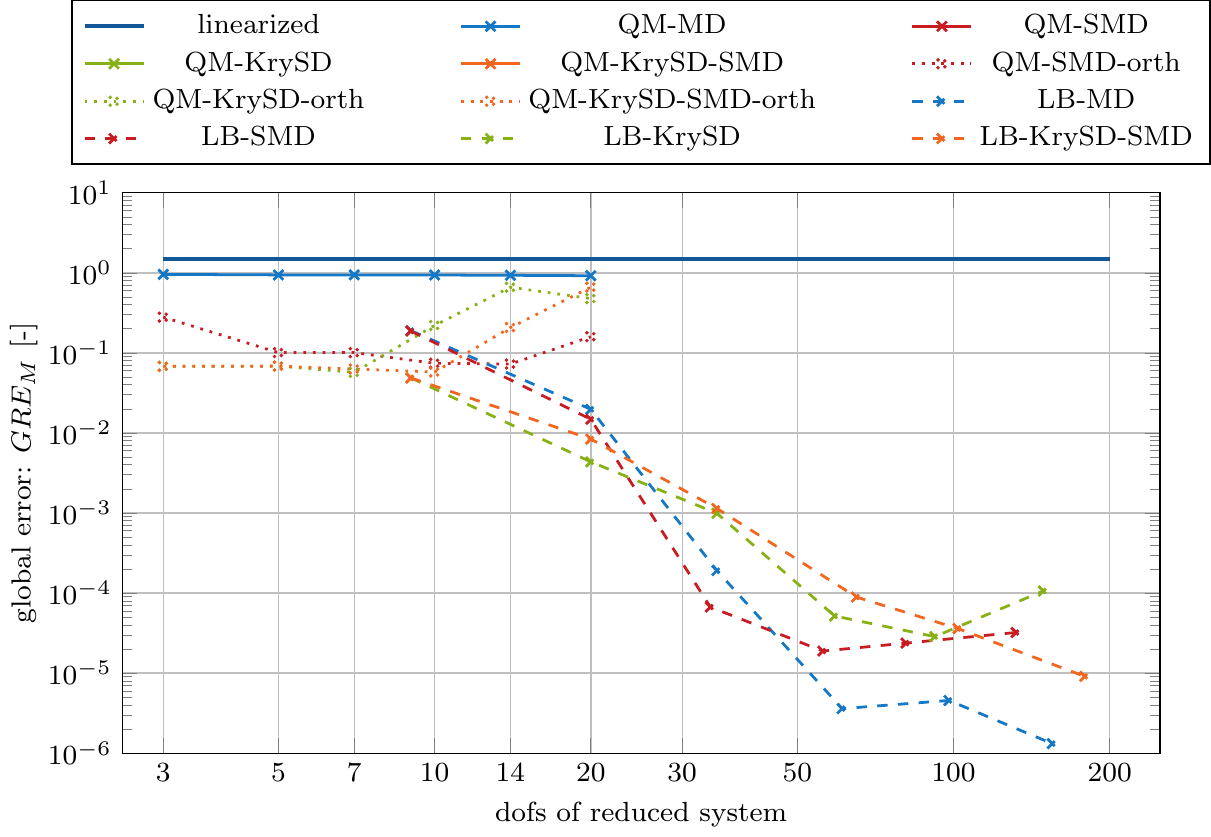}
   \caption{$GRE_M$ error of the benchmark example Arch for the different reduction techniques. } \label{fig:error_overview_arch}
\end{figure}

A plot of the displacement time series of a node in the middle of the arch is given in figure \ref{fig:time_evolution_arch}. In the case of 10 modes, all non-orthogonalized QM methods diverge except the QM-MD approach. However, this approach lacks accuracy. On the other hand, all orthogonalized QM approaches follow the reference line pretty well. A more extensive account of the $GRE_M$ errors with several reduction orders is shown in figure \ref{fig:error_overview_arch}. From this figure, it can be inferred that the QM-orth approach performs well, if a low number of dofs is adopted. However, as in the clamped-clamped beam example, the orthogonalization of $\vTheta$ strongly alters the quadratic manifold, when the number of modes increases.  This is reflected by a loss in accuracy. As in the previous examples, the best accuracy can be gained, when a LB method is used. This comes, however, at the price of many dofs.

\subsection{Cantilever}
\begin{figure}[htb]\centering
   \includegraphics[width=\picwidth]{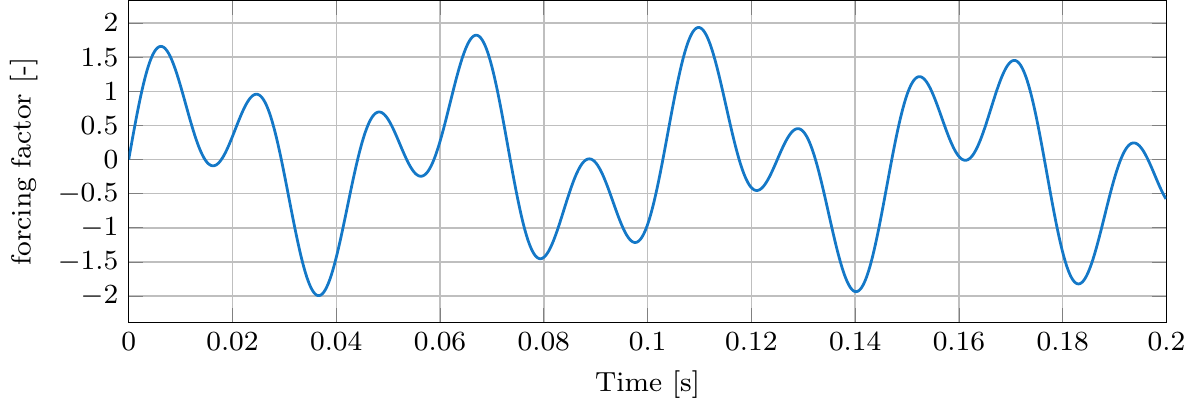}
   \caption{Time evolution of the forcing function of the Cantilever.} \label{fig:forcing_time_series_cantilever}
\end{figure}

The last of the investigated examples is a cantilever beam. The material parameters and the mesh are exactly the same as in the clamped-clamped beam example, but only with the left side clamped. The loading is applied on the right as a traction in vertical direction with an amplitude of $\unitfrac[3\cdot10^6]{N}{m}$ and the time series $g(t) = \sin(20\cdot2\pi t) + \sin(48\cdot2\pi t))$ also given in figure \ref{fig:forcing_time_series_cantilever}. The first eigenfrequencies of the linearized structure are $\omega_1 = \unitfrac[10.3\cdot2\pi]{rad}{s}$, $\omega_2 = \unitfrac[64.2\cdot2\pi]{rad}{s}$ and $\omega_3 = \unitfrac[179.1\cdot2\pi]{rad}{s}$.

\begin{figure}[htb]\centering
   \def\svgwidth{\picwidth}
   {\footnotesize
      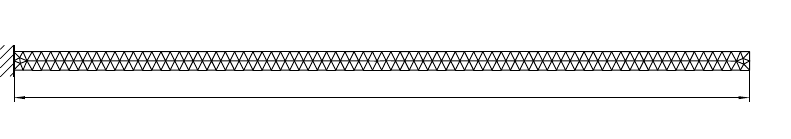}
   \caption{Mesh and boundary conditions of the cantilever example.}\label{fig:benchmark_cantilever}
\end{figure}

\begin{figure}[htb]\centering
   \includegraphics[width=\picwidth]{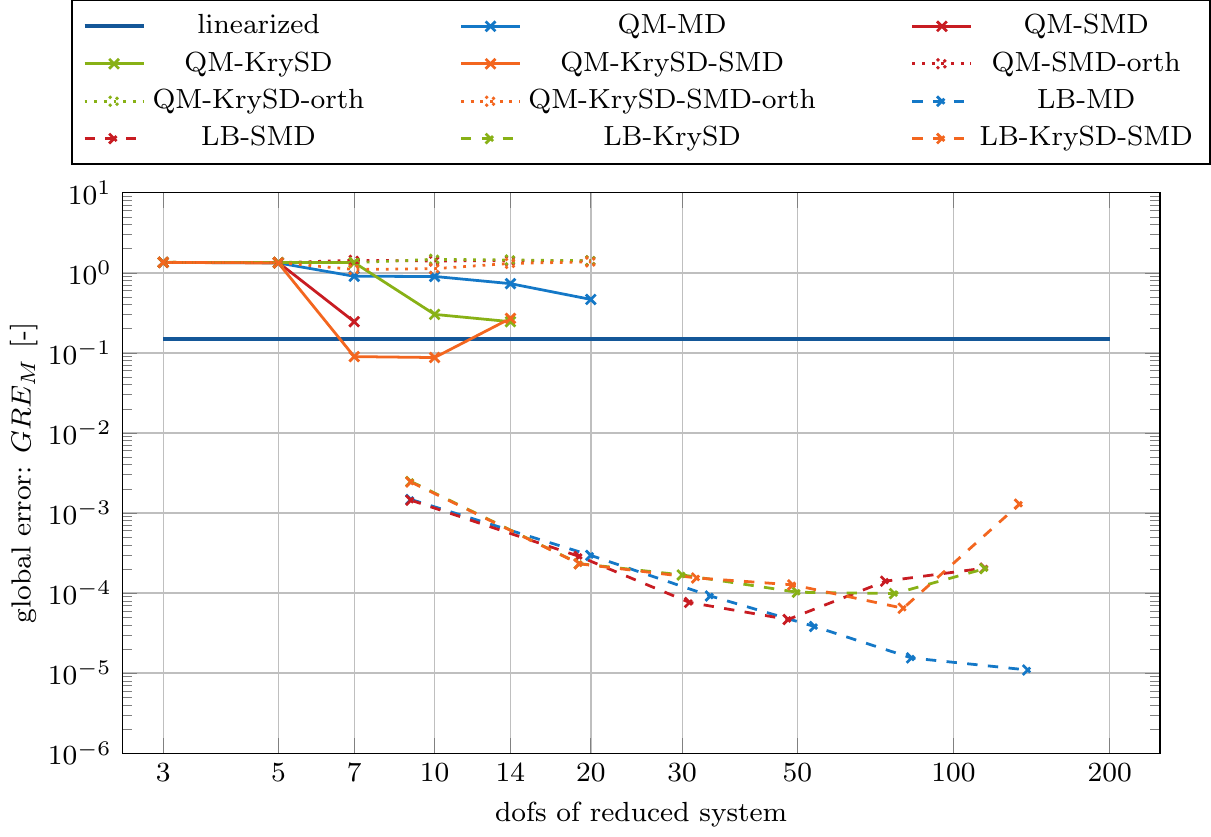}
   \caption{$GRE_M$ error of the benchmark example cantilever for the different reduction techniques. } \label{fig:error_overview_cantilever}
\end{figure}

Both the displacement time series of a point close to the tip of the cantilever in transverse direction depicted in figure~\ref{fig:forcing_time_series_cantilever} as well as the $GRE_M$ error for multiple reduction orders in figure~\ref{fig:error_overview_cantilever} show clearly, that the QM approach does not reproduce the dynamics of the original system. Furthermore, the linearized system shows a better accuracy than most of the QM approaches.

\begin{figure}[htb]\centering
   \includegraphics[width=\picwidth]{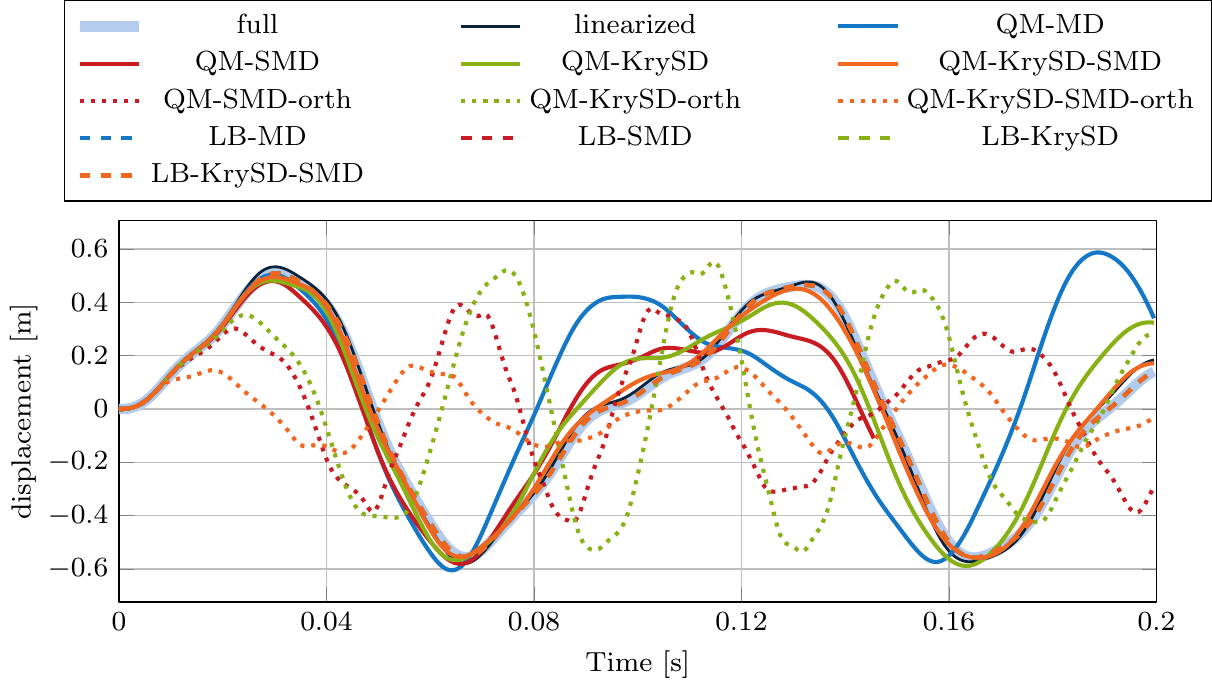}
   \caption{Time evolution of the vertical displacement of a point close to the tip. The reduction methods are run with 10 modes.} \label{fig:time_evolution_cantilever}
\end{figure}

On the other hand, the LB approach shows a good performance even for a relatively small number of dofs. Consequently, the subspace of the nonlinear motion is well captured by the (S)MDs, but the quadratic coupling within the QM does not represent the physical behavior.

\subsection{Discussion}

In the previous subsections, the QM approach with all its variants is investigated on four examples. They extend the examples of the companion paper \cite{tiso2016part_1}, as volume discretizations are used instead of shell elements with the von Karman kinematic assumptions. Some of the examples show good results for the QM reduction technique whereas others show a lack of accuracy or stability.
The instability of the QM approach can be eliminated by an orthogonalization of the quadratic term with respect to the linear terms, as proposed in remark~\ref{rem:orthogonalization_theta}. However, this orthogonalization destroys the QM structure and thus is only applicable to models with a small number of modes, where the violation of the QM relationship of modes and SDs is moderate, and thus good accuracy is gained.

In terms of accuracy, the examples showing good results for the QM approach are problem types where the so called \textit{cable effect} is the dominating nonlinear contribution. On the other hand, when large rotations become the relevant source of nonlinearity, the QM approach does not represent the physical behavior, as the quadratic coupling between modes and SDs is violated.

\begin{figure}[htb]\centering
   \includegraphics[width=\picwidth]{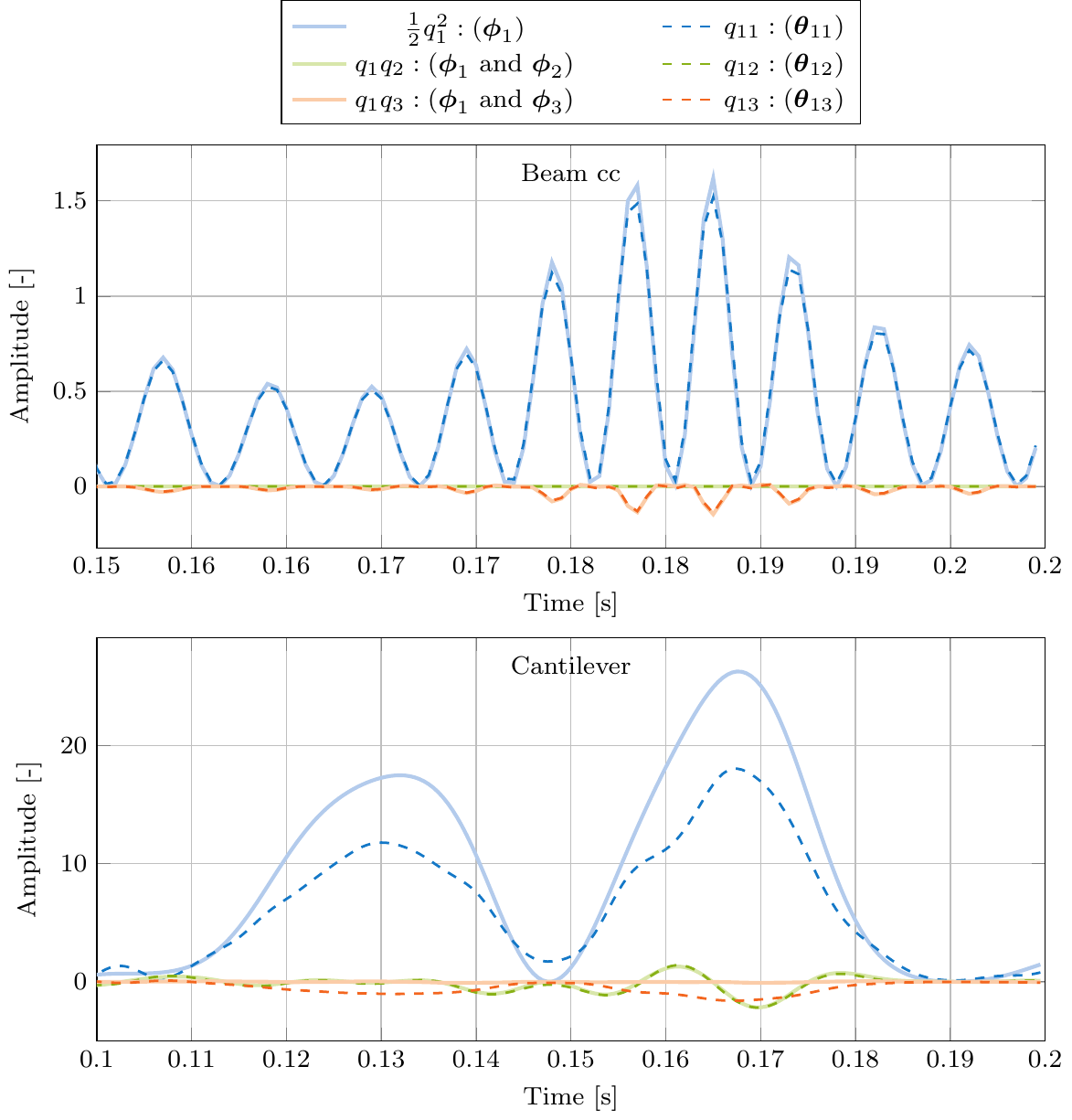}
   \caption{A posteriori comparison of the coupling between the squared modal amplitudes  and the amplitudes of the SMDs for the dominant modal contributions in the example Beam cc and Cantilever. The simulation was carried out with a linear basis containing 5 modes and all corresponding SMDs, so that modes and SMDs had their own dofs. \newline
   The top plot of example Beam cc shows a good match of the QM constraint, so that the squared modal amplitudes follow the amplitudes of the SMDs. The bottom plot of example Cantilever shows that the quadratic coupling of modes and SMDs within the QM gets violated. Note the different time scaling in the plot to show the time evolution in more detail for the higher frequency dynamics in example Beam cc.\newline
   The variables are as follows: $q_1$ is the amplitude of $\vphi_1$, $q_2$: amplitude of $\vphi_2$ and $q_3$: amplitude of $\vphi_3$, $q_{11}$: amplitude of $\vtheta_{11}$, $q_{12}$: amplitude of $\vtheta_{12}$, $q_{13}$: amplitude of $\vtheta_{13}$.} \label{fig:smd_coupling}
\end{figure}

One illustration of this effect is given in figure \ref{fig:smd_coupling}. The displacements of an LB run are projected onto the modes and the corresponding SMDs, so that the amplitudes associated to the modes and the SMDs can be reconstructed. They are compared to the amplitudes of the squared modal amplitudes as in a QM setting. In the example clamped-clamped beam, where the cable effect is dominating, the squared modal amplitudes coincide with the amplitudes of the SMDs, indicating that the QM coupling is a valid assumption.
In the cantilever-example, however, the squared modal amplitudes do not follow the amplitudes of the SMDs and consequently the QM coupling cannot represent the behavior of the cantilever beam. Here, we provide an explanation based on the von Karman finite element formulation.

The nonlinear restoring force of a geometrically nonlinear finite element formulation with a linear material constitutive law is a third order polynomial function, if $\vu$ is composed of nodal displacements only (and not  3D rotations).
For certain mechanical configurations with kinematic assumptions like a straight beam, however, the dofs are not fully polynomially coupled.

For a straight beam using von Karman kinematics, the Green-Lagrange strain $\varepsilon_x$ is given as \cite{crisfield2012nonlinear, rutzmoser2014model}
\begin{align}
\varepsilon_{x} = \frac{\d u}{\d x} - z \frac{\d^2 w}{\d x^2} + \frac{1}{2}\left( \frac{\d w}{\d x} \right)^2.
\end{align}
with the displacements $u$ and $w$ in $x$ and $z$ direction, respectively. The $x$ axis goes along the beam axis and the $z$ axis goes perpendicular, starting from the centroid of the cross section.

If a linear constitutive law is used, the finite element approximation results in the following equation of motion in $x$ and $z$ direction indicated with $m$ for membrane and $b$ for bending:
\begin{gather}
M_{m,ij}\;\ddot{u}_j + K^{(1)}_{m,ij}\;u_j + K^{(2)}_{mb,ijk}\;w_j\;w_k = F_{m,i}\,, \label{eq:membrane_general}\\
M_{b,ij}\;\ddot{w}_j + K^{(1)}_{b,ij}\;w_j + K^{(2)}_{bm,ijk}\;w_j\;u_k + K^{(3)}_{b,ijkl}\;w_j\;w_k\;w_l = F_{b,i}\,, \label{eq:bending_general}
\end{gather}
with the mass matrices $M_m$ and $M_b$, the linear stiffness matrices $K^{(1)}_m$ and $K^{(1)}_b$, the third order stiffness tensors $K^{(2)}_{mb}$ and $K^{(2)}_{bm}$ generating the quadratic force contributions, and the fourth order stiffness tensor $K^{(3)}_b$ generating the cubic forces.

The two sets of equations above state that, besides the internal dynamics, the axial displacements $u$ are triggered by the squares of the transverse displacements~$w$. If the axial dynamics happen at a faster time scale compared to the transverse ones (as can be expected of a beam), one can neglect the inertial terms in \eqref{eq:membrane_general} to write the axial displacements $u$ as a quadratic function of the transverse dispclacements $w$ as
\begin{align}
u_j = - (K_{m}^{(1)^{-1}})_{ji} K^{(2)}_{mb,ikl}\;w_k\;w_l \label{eq:static_condensation}\,.
\end{align}
This leads to the quadratic manifold constructed via the force compensation method, where the basis $\vV$ contains all the transverse displacement $w$ but not the axial ones $u$ (cf. \cite{rutzmoser2014model} and \cite{rutzmoser2014isma}). Thus, if the basis $\vV$ contains solely transverse modes, the QM statically condenses the axial displacements as in \eqref{eq:static_condensation}. 

Consequently, the quadratic manifold obtained using the force compensation technique is an appropriate reduction technique, when a decoupling as in the von Karman assumption above is possible, and the dynamics in the condensed direction of the QM are relatively fast. The decoupling can either be enforced explicitly in the finite element formulation, as is the case when von Karman shells or beams are used (cf. companion paper \cite{tiso2016part_1}), or it is implicitly given with the mechanical problem independent of the FE formulation, as in the examples of clamped-clamped Beam and Hat given in this paper. Both examples have in common the cable or the membrane effect as the dominating source of nonlinearity.

If the structure of the nonlinearity can not be decomposed as explained above, the QM approach can lead to poor results, as the cantilever beam example shows. For these types of systems, the von Karman assumption is not valid and both, axial and transverse displacements have fully coupled third order polynomial restoring forces. As a result, the QM approach imposes unphysical constraints on the system, leading to an undesirable stiffening or locking effect. However, the (S)MDs and SDs still capture the subspace of the nonlinear motion fairly well making them an attractive constituent for LB approaches.

\section{Conclusion}
In this work a Quadratic Manifold (QM) approach for the reduction of geometrically nonlinear structural dynamic systems is extended to arbitrary linear modes equipped with a sound theory called the \emph{Force Compensation Approach}. The Static Derivatives (SDs) forming the quadratic part in the extended QM framework can be interpreted as the natural extension of the Static Modal Derivatives (SMDs) for arbitrary linear modes.

While in the companion paper \cite{tiso2016part_1} the feasibility of the approach is demonstrated with shell elements incorporating the von Karman kinematic assumption which leads to a natural decoupling of modes and corresponding (Static) Modal Derivatives, it is shown here in a numerical study of several examples, that the QM approach is also an attractive reduction method independent of the type of finite element discretization. 

When the so called cable effect is the dominating source of nonlinearity, the QM approach exhibits high accuracy while keeping the number of unknowns minimal.  For problem types comprising large rotations as predominant nonlinearity, the QM approach leads to poor results. The SDs, however, perform remarkably well in a linear basis without the constraint of a quadratic coupling. 

The extension of the QM approach to arbitrary linear bases is advantageous, as the accuracy of the QM approach can be enhanced in many cases by substituting the vibration modes forming the linear basis with Krylov subspace vectors.

\section*{References}
\bibliographystyle{plain}
\bibliography{literature}

\end{document}